\documentstyle [12pt,a4p,epsfig,amsmath,multicol]{article}
\newcommand{\Cp}{{\rm C}'}
\newcommand{\Cpp}{{\rm C}''}

\begin{document}
\vspace*{0.6cm}

\begin{center} 
{\normalsize\bf The Hafele-Keating experiment, velocity and length interval
 transformations and resolution of the Ehrenfest paradox}
\end{center}
\vspace*{0.6cm}
\centerline{\footnotesize J.H.Field}
\baselineskip=13pt
\centerline{\footnotesize\it D\'{e}partement de Physique Nucl\'{e}aire et 
 Corpusculaire, Universit\'{e} de Gen\`{e}ve}
\baselineskip=12pt
\centerline{\footnotesize\it 24, quai Ernest-Ansermet CH-1211Gen\`{e}ve 4. }
\centerline{\footnotesize E-mail: john.field@cern.ch}
\baselineskip=13pt
\vspace*{0.9cm}
\abstract{ A relativistic analysis based on the paths, in a non-rotating frame comoving with the centroid of 
 the Earth, of clocks carried by aircraft circumnavigating the Earth in different directions, as in
 the Hafele-Keating experiment, predicts time differences between airborne and Earth-bound clocks
 at variance with the results of the experiment. The latter imply new relativistic velocity
 transformations differing from the conventional ones. These transformations demonstrate in turn
 the invariance of length intervals on the surface of the rotating Earth and so resolve
 the Ehrenfest paradox for this case. }
\vspace*{0.9cm}
\normalsize\baselineskip=15pt
\setcounter{footnote}{0}
\renewcommand{\thefootnote}{\alph{footnote}}

  In the Hafele-Keating (HK) experiment~\cite{HK}, performed in 1971, four caesium-beam atomic clocks
  were flown around the world in commercial aircraft, once in the west-to-east (W$-$E) and once in the
  east-to-west (E$-$W) direction. The time intervals recorded by the clocks during the flights were
  compared with those recorded by reference clocks at the U.S. Naval Observatory. The time intervals
  for the airborne clocks were sensitive to both special relativitistic (SR) and general relativistic (GR), or
  gravitational,
  effects. Here only SR effects are considered. It is assumed, for simplicity, in the following calculations,
   that the clock
  $\Cpp$ at rest in the comoving frame S'' of the aircraft executes an equatorial circumnavigation 
  of the Earth at constant speed $v'_A$ in the comoving frame S' of the ground-based clock $\Cp$. 
   The latter moves with constant speed $v_E = \Omega R$ relative to a non-rotating inertial frame S comoving
 with the centroid of the Earth; that is, the rotation of the Earth around the Sun is neglected.
  The parameter $\Omega$ is the angular frequency of rotation of the Earth and $R$ is its equatorial 
  radius. Since gravitational effects are neglected, the altitude of the aircraft during the flights
   may be neglected in comparision with $R$.
  \par The clocks  $\Cp$ and $\Cpp$ undergo transverse acceleration due to the rotation of the Earth, but
   experiments with decaying muons in near-circular orbits in a storage ring at CERN demonstrated that the
   special-relativistic time dilation (TD) effect is the same as for uniform motion at the same speed, $v$, where
   $v/c = 0.9994$ or $\gamma = \sqrt{1-(v/c)^2} = 29$, with a relative precision of $0.1$ $\%$ in the presence 
    of a transverse acceleratation, due to the bending field of the storage ring, of $10^{18}g$~ \cite{CERNTD}.
     The time dilation and other relativistic
    effects in the HK experiment can therefore be calculated with confidence on the assumption that S'' and S' are inertial
    frames moving with speeds $v'_A$ and $v_E$ relative to the frames S' and S respectively.

    \par For the analysis  of the experiment a hypothetical clock, C, at rest in the frame S, registering
    `coordinate time' is introduced~\cite{HN,HAJP}. If $ T'$ is the time interval recorded by the Earth-bound
     clock during either the W$-$E or the E$-$W flights, then
                \begin{equation}
           T' = \frac{2 \pi R}{ v'_{{\rm A}}}.
            \end{equation}
     If the distances and times travelled by the aircraft in the frame S during the round trips are denoted 
     by $d$, $T$ respectively (where $T$ denotes an unobserved coordinate time interval registered by C)
     then:
             \begin{eqnarray}
         d({\rm W-E}) & = &  v_{{\rm A}}({\rm W-E})T({\rm W-E}) = v_{{\rm E}}T({\rm W-E})+2 \pi R, \\ 
          d({\rm E-W}) & = & v_{{\rm A}}({\rm E-W})T({\rm E-W}) = v_{{\rm E}}T({\rm E-W})-2 \pi R.                            
        \end{eqnarray}
       Using the conventional relativistic parallel velocity addition relations to give the velocity
     $\hat{v}_{{\rm A}}$ of the aircraft in the frame S:
      \begin{eqnarray}
       \hat{v}_{{\rm A}}({\rm W-E}) & = & \frac{v_{{\rm E}} + v'_{{\rm A}}}{1+\frac{v_{{\rm E}} v'_{{\rm A}}}{c^ 2}},
           \\
  \hat{v}_{{\rm A}}({\rm E-W}) & = & \frac{ v_{{\rm E}} - v'_{{\rm A}}}{1-\frac{ v_{{\rm E}} v'_{{\rm A}}}{c^ 2}}.       
       \end{eqnarray}
       Setting $ v_{{\rm A}}=\hat{v}_{{\rm A}}$ in (2) and (3), and using Eq.~(1) to eliminate $2 \pi R$, the flight times in the frame S are found to be:
      \begin{eqnarray}
       \hat{T}({\rm W-E}) & \equiv & \frac{d({\rm W-E})}{\hat{v}_{{\rm A}}({\rm W-E})}
        =   T'\gamma(\Cp)^2(1+\frac{ v_{{\rm E}} v'_{{\rm A}}}{c^ 2}),  \\
      \hat{T}({\rm E-W}) & \equiv & \frac{d({\rm E-W})}{\hat{v}_{{\rm A}}({\rm E-W})}
        =   T'\gamma(\Cp)^2(1-\frac{v_{{\rm E}} v'_{{\rm A}}}{c^ 2})
   \end{eqnarray}
      where  $\gamma(\Cp) \equiv 1/\sqrt{1-(v_{{\rm E}}/c)^2}$.
     \par  The TD effect between the frames S and S'' is:
       \begin{equation}
        \Delta t = \gamma(\Cpp)  \Delta t''  
        \end{equation}
      where
     \begin{equation}
      \gamma(\Cpp) =  \gamma'(\Cpp) \gamma(\Cp)(1\pm\frac{v_{{\rm E}} v'_{{\rm A}}}{c^ 2})
         \end{equation}
         and  $\gamma'(\Cpp) \equiv 1/\sqrt{1-(v'_{{\rm A}}/c)^2}$. In Eq.~(9), the $+ (-)$ signs correspond to 
          the W$-$E (E$-$W) flights. The formula (9) is the Lorentz transformation of the TD factor $\gamma$ for the clock 
          $\Cpp$ between the frames S' and S. It is algebraically equivalent to Eqs.~(4) and (5). The TD factor $\gamma$
        is the temporal component of the dimensionless 4-vector velocity of the clock and obeys the 4-vector
       Lorentz transformation formula, of which (9) is an example. 
        \par The time difference observed for the W$-$E flight is
  \begin{equation}
          \Delta \hat{T}'({\rm W-E}) \equiv T''({\rm W-E})-T' =T'\left(\frac{ T''({\rm W-E})}{T'}-1\right).
         \end{equation}
        Setting $ \Delta t =  \hat{T}({\rm W-E})$ and  $ \Delta t''  = T''({\rm W-E})$ in (8) 
         and eliminating the unmeasured coordinate time interval $\hat{T}({\rm W-E})$ between the resulting equation
         and (6) gives:
        \begin{equation}
           \frac{ T''({\rm W-E})}{T'} = \frac{\gamma(\Cp)^2}{\gamma(\Cpp)}(1+\frac{ v_{{\rm E}} v'_{{\rm A}}}{c^ 2}) 
             =   \frac{\gamma(\Cp)}{\gamma'(\Cpp)}
        \end{equation}
           where in the last member (9) has been used to eliminate $\gamma(\Cpp)$. 
          A similar calculation for the E$-$W flight shows that 
       \begin{equation}
           \frac{ T''({\rm E-W})}{T'} = \frac{\gamma(\Cp)}{\gamma'(\Cpp)} =  \frac{ T''({\rm W-E})}{T'}.
        \end{equation}
         Therefore Eqs.~(10)-(12) give:
         \begin{equation}
          \Delta \hat{T}'({\rm W-E}) = \Delta \hat{T}'({\rm E-W})
              = T'\left(\frac{\gamma(\Cp)}{\gamma'(\Cpp)}-1\right).
          \end{equation}
           Equal time differences are therefore predicted for the W$-$E and the E$-$W flights. The actual parameters
           of the HK experiment are well approximated by the constant values: $ v'_{{\rm A}} = 300$m/s, 
           $v_{{\rm E}} = \Omega R = 470$m/s and $T' = 37.1$h, for which (13) predicts:   
         \[ \Delta \hat{T}'({\rm W-E}) = \Delta \hat{T}'({\rm E-W}) =  97{\rm ns} \]      
       Which may be compared with the predictions for the special-relativistic (SR) effect in the actual
       HK experiment derived by properly taking into account the actual paths followed by the aircraft
       over the Earth's surface as well as the time-dependence of their speeds~\cite{HK}: 
    \[  \Delta T'_{{\rm HK}}({\rm W-E})_{{\rm SR}} = -184 \pm 18{\rm~ ns} \]
    \[  \Delta T'_{{\rm HK}}({\rm E-W})_{{\rm SR}} = 96  \pm 10{\rm~ ns} \] 
 Including GR effects the overall prediction for the time differences was~\cite{HK} 
\[  \Delta T'_{{\rm HK}}({\rm W-E})_{{\rm SR+GR}} = -40 \pm 23{\rm~ ns} \]
    \[  \Delta T'_{{\rm HK}}({\rm E-W})_{{\rm SR+GR}} = 275  \pm 21{\rm~ ns} \]
     which were found to be in good agreement with experimentally measured  values~\cite{HK}:
  \[  \Delta T'_{{\rm HK}}({\rm W-E})_{{\rm meas}} = -59 \pm 10{\rm~ ns} \]
    \[  \Delta T'_{{\rm HK}}({\rm E-W})_{{\rm meas}} = 273  \pm 7{\rm~ ns} \]
          Calculating instead the overall prediction by replacing the SR predictions of Ref.~\cite{HK}
         with those given by Eq.~(13) gives
\[  \Delta \hat{T}'_{{\rm HK}}({\rm W-E})_{{\rm SR+GR}} = 141{\rm~ ns} \]
    \[  \Delta \hat{T}'_{{\rm HK}}({\rm E-W})_{{\rm SR+GR}} = 282{\rm~ ns} \]
   incompatible with $\Delta T'_{{\rm HK}}({\rm W-E})_{{\rm meas}}$. 
     \par It is then clear that the calculation above, based on Eqs.~(1)-(5), (8) and (9) does not describe correctly
     the results of the HK experiment. On reflection it is quickly seen that the mistake resides
     not in the purely geometrical formulae (2) and (3), the TD relation (8) or the transformation
     law (9), but in the velocity transformation formulae
     (4) and (5). These predict, when inserted in (2) and (3), that
     $\hat{T}({\rm W-E}) \ne \hat{T}({\rm E-W})$. Suppose that the aircraft start out at the same instant,
      and travel with the same speed relative to the surface of the Earth during the W$-$E and E$-$W flights.
       They will arrive back simultaneously at their starting point. There is thus a triple
      world line coincidence event (those of the two aircaft and the starting point on the Earth) at
      arrival. This must be observed as such in all frames, including S. It is therefore impossible,
       by this `zeroth theorem of space-time physics', the importance of which has previously been 
        stressed by Langevin~\cite{ Langevin,LangJHF} and Mermin~\cite{Mermin}, that 
        $T({\rm W-E}) \ne T({\rm E-W})$. The velocity transformation formulae (4) and (5) are therefore
        inapplicable to the analysis of the HK experiment in the way shown above.
         \par Indeed, the correct SR prediction for the HK experiment can be obtained without any
         consideration of the distances $d({\rm W-E})$ and $d({\rm E-W})$ covered by the aircraft 
         in the frame S during the round trips. The TD effect between the frames S and S' is given
        by the relation
        \begin{equation}
         \Delta t = \gamma(\Cp) \Delta t' 
        \end{equation}
        from which it is clear (contrary to Eqs.(6) and (7)) that  $T({\rm W-E}) = T({\rm E-W})$.
        It is then found by combining Eqs.~(8),(9),(10) and (14) that
          \begin{equation}
          \Delta T'({\rm W-E}) =T'\left(\frac{\gamma(\Cp)}{\gamma(\Cpp)}-1\right)
                = T'\left(\frac{1}{\gamma'(\Cpp)\left[1+\frac{v_{{\rm E}} v'_{{\rm A}}}{c^ 2}\right]}-1\right).
   \end{equation}
         \par Retaining only O($\beta^2$) terms in (15) and the corresponding formula for
          $\Delta T'({\rm E-W})$ gives
       \begin{eqnarray}
       \Delta T'({\rm W-E})& =  & -\frac{T'\beta'_{{\rm A}}}{2}( \beta'_{{\rm A}}+ 2 \beta_{{\rm E}}), \\ 
      \Delta T'({\rm E-W})& =  & \frac{T'\beta'_{{\rm A}}}{2}( -\beta'_{{\rm A}}+ 2 \beta_{{\rm E}})
   \end{eqnarray}
        where $\beta'_{{\rm A}} \equiv v'_{{\rm A}}/c$, $\beta_{{\rm E}} \equiv v_{{\rm E}}/c$.
         Substituting the numerical values of $ v'_{{\rm A}}$ and  $v_{{\rm E}}$ quoted above in (16) and (17)
         gives
    \[  \Delta T'_{{\rm HK}}({\rm W-E})_{{\rm SR}} = -276{\rm~ ns} \]
    \[  \Delta T'_{{\rm HK}}({\rm E-W})_{{\rm SR}} = 143{\rm~ ns} \]
     in qualitative agreement with the calculated predictions for the HK experiment. 
     \par Setting $T({\rm W-E})= T({\rm E-W}) = T = \Delta t$, $T' = \Delta t'$ and using 
      Eqs.~(1),(2),(3) and (14), the correct velocity 
         transformation formulae, between the frames S' and S, for the aircraft are
         \begin{eqnarray}
     v_{{\rm A}}({\rm W-E}) &\equiv& \frac{d({\rm W-E})}{T} =  v_{{\rm E}}+\frac{v'_{{\rm A}}}{\gamma(\Cp)}, \\
     v_{{\rm A}}({\rm E-W}) &\equiv& \frac{d({\rm E-W})}{T} =  v_{{\rm E}}-\frac{v'_{{\rm A}}}{\gamma(\Cp)}.
         \end{eqnarray}
        These transformation formulae for relative velocities between different inertial frames in
        the same space-time experiment  were obtained, in terms of corresponding angular
    velocities, in a review article on the Sagnac effect~\cite{Post} published by Post in 1967 and have been previously derived by the present author~\cite{JHFSTP3,JHFRECP}. Similar formulae were also previously considered by Selleri~\cite{Selleri}
    and Klauber~\cite{Klauber}.
        \par Consider now the distance, $\Delta s'({\rm W-E})$, in the frame S', of the aircraft from its starting
      point after a time interval, $\Delta t'$, sufficiently short that the curvature of the surface of the Earth may be 
      neglected. Denoting by $\Delta d({\rm W-E})$ and $\Delta t$ the corresponding distance moved and time interval, in the frame S,
     then (18) and (14) give:
         \begin{eqnarray}
     \Delta d({\rm W-E})&  =  &  v_{{\rm A}}\Delta t =  v_{{\rm E}}\Delta t + \frac{v'_{{\rm A}}\Delta t} {\gamma(\Cp)} 
       \nonumber \\         
        &  =  &  v_{{\rm E}}\Delta t + v'_{{\rm E}} \Delta t' \nonumber \\ 
         &  =  &  v_{{\rm E}}\Delta t+\Delta s'({\rm W-E}).
        \end{eqnarray}
        Denoting by $\Delta s({\rm W-E})$ the distance in S between the aircraft and the starting point of 
        its flight after the time interval $\Delta t$, transposition  of (20) gives:
          \begin{equation}
        \Delta s({\rm W-E}) = \Delta d({\rm W-E})-  v_{{\rm E}}\Delta t = \Delta s'({\rm W-E}).
          \end{equation}
          The length interval between the aircraft and its starting point is therefore the same in the
          frames S and S' in relative motion ---there is no `length contraction' effect.
         \par This demonstration resolves the Ehrenfest paradox~\cite{Ehrenfest} concerning the ratio of
        the circumference to the radius of a rotating disc. This ratio is simply $2 \pi$. Contrary to
         Einstein's assertions~\cite{EinE1,EinE2}, the ratio is not greater than  $2 \pi$ and no introduction
        of non-Euclidean spatial geometry is necessary.
     \par The conventional relativistic transformation  formulae (4) and (5) are not incorrect, but only misinterpreted
         in the calculation above. These formulae do correctly describe a kinematical transformation between
        configurations of two different and {\it physically independent} space time
         experiments~\cite{JHFSTP3,JHFRECP}, not velocities as observed in different frames of { \it the
         same} space time experiment, as assumed above. In fact, writing $\hat{\gamma}(\Cpp) = 1/\sqrt{1-(\hat{v}_{{\rm A}}/c)^2}$
         it may be shown that $\hat{\gamma}(\Cpp) = \gamma(\Cpp)$ so that the velocity transformations equations (4) and (5)
      are algebraically equivalent to Eq.~(9),
          which is the transformation equation of the TD factor $\gamma$ for the clock $\Cpp$ between the frames S' and S.
           \par In conclusion, the experimental results of the HK experiment falsify
              the conventional interpretation of the relativistic velocity transformation formulae
              (4) and (5), since the latter, when used to calculate flight times in the frame S
              predict equal time differences between the airborne and Earth-bound clocks for the W$-$E and E$-$W flights.
             The necessary equality of the S-frame durations of the W$-$E and E$-$W flights (in contradiction
                 with the predictions, (6) and (7), of (4) and (5) respectively) requires
               the velocity transformation formulae for the HK experiment, between the frames S' and S,
               to be (18) and (19). These equations show further that there is no `length contraction' effect for
               spatial intervals on the surface of the Earth
               and so resolve the corresponding Eherenfest paradox for the radius and equatorial
               circumference of the rotating Earth. How the spurious and correlated `length contraction'
                 and `relativity of simultaneity' effects of conventional special relativity
                arise from a general and fundamental misinterpretation of the space-time Lorentz transformation
              is explained
                 elsewhere~\cite{JHFUMC,JHFCRCS,JHFACOORD,JHFFJMP1,JHFFJMP2}.
             \par{\bf Acknowledgement}
               I am indepted to Brian Coleman for sending me a draft of
               his paper on the HK experiment containing the important S-frame path equations (2) and (3),
               which I had not previously noticed.

\end{document}